\documentclass[sigconf]{acmart}
\usepackage{xcolor}
\usepackage{multirow}
\usepackage{makecell}
\usepackage{hyperref}
\usepackage{cleveref}
\usepackage{colortbl}
\usepackage{tcolorbox}
\tcbuselibrary{breakable}
\usepackage{siunitx}
\usepackage{xspace}
\usepackage{tikz}
\usepackage{enumitem}
\usetikzlibrary{arrows}
\usetikzlibrary{arrows.meta}
\usetikzlibrary{patterns}
\usepackage{pgfplots}
\usepgfplotslibrary{groupplots}
\usepgfplotslibrary{statistics}
\usetikzlibrary{calc}
\usepackage{balance}

\definecolor{color-a}{RGB}{244, 241, 222}
\definecolor{color-b}{RGB}{129, 178, 154}
\definecolor{color-c}{RGB}{61, 64, 91}
\definecolor{color-d}{RGB}{242, 204, 143}
\definecolor{color-e}{RGB}{224, 122, 95}
\definecolor{color-f}{RGB}{201, 228, 202}
\definecolor{color-g}{RGB}{254, 217, 183}

\definecolor{gray}{gray}{0.9}
\definecolor{light-gray}{gray}{0.4}

\definecolor{lightgray}{RGB}{248,249,250}
\definecolor{level1}{RGB}{242,248,245}
\definecolor{level2}{RGB}{216,232,222}
\definecolor{level3}{RGB}{191,215,199}
\definecolor{level4}{RGB}{160,196,176}
\definecolor{level5}{RGB}{129,178,154}
\definecolor{level6}{RGB}{103,150,126}
\definecolor{level7}{RGB}{77,123,99}
\definecolor{level8}{RGB}{51,95,71}

\newcommand{\benchmark}{\textsc{BenchVul}\xspace}
\newcommand{\dataset}{\textsc{TitanVul}\xspace}
\newcommand{\rvg}{\textsc{RVG}\xspace}

\newcommand{\myparagraph}[1]{\vspace{4px}\noindent{\textbf{#1}}\hspace{6pt}}
\newcommand{\takeaway}[1]{
\begin{tcolorbox}[
  leftrule=0.5mm, rightrule=0.5mm, toprule=0.5mm, bottomrule=0.5mm,
  left=2pt, right=2pt, top=2pt, bottom=2pt,
  breakable
  ]%
\em #1
\end{tcolorbox}
}

\newcolumntype{L}[1]{>{\raggedright\let\newline\\\arraybackslash\hspace{0pt}}m{#1}}
\newcolumntype{C}[1]{>{\centering\let\newline\\\arraybackslash\hspace{0pt}}m{#1}}
\newcolumntype{R}[1]{>{\raggedleft\let\newline\\\arraybackslash\hspace{0pt}}m{#1}}

\AtBeginDocument{%
  }

\setcopyright{none}
\renewcommand\footnotetextcopyrightpermission[1]{}

\copyrightyear{2026}
\acmYear{2026}
\setcopyright{cc}
\setcctype{by}
\acmConference[ICSE '26]{2026 IEEE/ACM 48th International Conference on Software Engineering}{April 12--18, 2026}{Rio de Janeiro, Brazil}
\acmBooktitle{2026 IEEE/ACM 48th International Conference on Software Engineering (ICSE '26), April 12--18, 2026, Rio de Janeiro, Brazil}
\acmPrice{}
\acmDOI{10.1145/3744916.3787830}
\acmISBN{979-8-4007-2025-3/2026/04}

\begin{document}

\title{Out of Distribution, Out of Luck: How Well Can LLMs Trained on Vulnerability Datasets Detect Top 25 CWE Weaknesses?}

\author{%
Yikun Li$^{*}$, 
Ngoc Tan Bui$^{*}$, 
Ting Zhang\textsuperscript{$\heartsuit$}\textsuperscript{$\diamondsuit$}, 
Chengran Yang$^{*}$, 
Xin Zhou$^{*}$, 
Martin Weyssow$^{*}$, 
Jinfeng Jiang$^{*}$, 
Junkai Chen$^{*}$, 
Huihui Huang$^{*}$, 
Huu Hung Nguyen$^{*}$, 
Chiok Yew Ho$^{\S}$, 
Jie Tan$^{\ddagger}$, 
Ruiyin Li$^{\P}$, 
Yide Yin$^{\dagger}$, 
Han Wei Ang$^{\dagger}$, 
Frank Liauw$^{\dagger}$, 
Eng Lieh Ouh$^{*}$, 
Lwin Khin Shar$^{*}$, 
David Lo$^{*}$%
}
\thanks{$\diamondsuit$ Ting Zhang is the corresponding author (ting.zhang@monash.edu).}

\affiliation{%
  \institution{$^{*}$Singapore Management University, Singapore}\country{}
}
\affiliation{%
  \institution{\textsuperscript{$\heartsuit$} Monash University, Australia}\country{}
}
\affiliation{%
  \institution{$^{\S}$Chinese University of Hong Kong, China}\country{}
}
\affiliation{%
  \institution{$^{\ddagger}$University of Groningen, The Netherlands}\country{}
}
\affiliation{%
  \institution{$^{\P}$Wuhan University, China}\country{}
}
\affiliation{%
  \institution{$^{\dagger}$GovTech, Singapore}\country{}
}

\renewcommand{\shortauthors}{Li et al.}

\begin{abstract}
Automated vulnerability detection research has made substantial progress, yet its real-world impact remains limited. Prior work found that current vulnerability datasets suffer from issues including label inaccuracy rates of 20\%-71\%, extensive duplication, and poor coverage of critical Common Weakness Enumeration (CWE). These issues create a significant ``generalization gap'' where models achieve misleading In-Distribution (ID) accuracies (testing on splits from the same dataset) by exploiting spurious correlations rather than learning true vulnerability patterns.

To address these limitations, we present a three-part solution. First, we introduce \benchmark, which is a manually curated and balanced \emph{test dataset} covering the MITRE Top 25 Most Dangerous CWEs, to enable fair model evaluation. Second, we construct a high-quality \emph{training dataset}, \dataset, comprising 38,548 functions by aggregating seven public sources and applying deduplication and validation using a novel multi-agent LLM pipeline. Third, we propose a Realistic Vulnerability Generation (\rvg) pipeline, which synthesizes context-aware vulnerability examples for underrepresented but critical CWE types through simulated development workflows.

Our evaluation reveals that In-Distribution (ID) performance does not reliably predict Out-of-Distribution (OOD) performance on \benchmark. For example, a model trained on BigVul achieves the highest 0.703 ID accuracy but fails on \benchmark's real-world samples (0.493 OOD accuracy). Conversely, a model trained on our \dataset achieves the highest OOD performance on both the real-world (0.881) and synthesized (0.785) portions of \benchmark, improving upon the next-best performing dataset by 5.3\% and 11.8\% respectively, despite a modest ID score (0.590). Augmenting \dataset with our \rvg further boosts this leading OOD performance, improving accuracy on real-world data by 5.8\% (to 0.932).
Data is available at: \url{https://github.com/yikun-li/TitanVul-BenchVul}.
\end{abstract}



\maketitle

\section{Introduction}

Automated vulnerability detection is a popular area of software engineering research \cite{chakraborty2021deep,ding2024vulnerability,gao2023far,zhou2024large,zhang2025benchmarking,zhou2025large,lo2023trustworthy}. A survey reported that 88\% of studies in machine learning for vulnerability detection (ML4VD) approach the problem as function-level classification: given a function's source code, the task is to determine whether it contains a vulnerability \cite{risse2025top, liu2024vuldetectbench, croft2023data}. However, prior work has identified significant data quality issues in widely used vulnerability datasets, including high rates of label inaccuracy (20\%-71\%) and extensive data duplication \cite{risse2025top, liu2024vuldetectbench, croft2023data,ding2024vulnerability,chen2023diversevul,li2024cleanvul,weyssow2025r2vul}. In addition, as shown in \Cref{sec:empirical_study}, available datasets are imbalanced and often contain incorrect or outdated CWE labels.

Such dataset issues inflate model performance when evaluations rely on In-Distribution (ID) settings, which assess test splits drawn from the same dataset used in training. Inflated performance arises from models capturing dataset-specific biases rather than learning genuine vulnerability patterns \cite{risse2025top}, causing a gap between ID performance and real-world effectiveness, which requires Out-of-Distribution (OOD) evaluation on independent, unseen data. We define this difference between ID and OOD performance as the generalization gap. This gap undermines reliable model comparisons and assessments of dataset quality. We notice three challenges:

\myparagraph{Challenge I: Unreliable Evaluation Due to Overfitting}
Currently, researchers rely primarily on ID evaluations that use test samples drawn from the same datasets used for training. As these datasets often contain duplication and labeling issues, models may achieve high ID performance by memorizing dataset-specific artifacts rather than learning generalizable patterns. Consequently, the actual capability of models remains uncertain \cite{risse2025top}.

\myparagraph{Challenge II: Poor-Quality Training Data at Scale}
Widely-used vulnerability datasets suffer from low data quality, including high rates of noise, irrelevant code changes, refactoring, and non-security-related fixes \cite{croft2023data,risse2025top}. This limitation prevents models from learning robust and generalizable vulnerability patterns.

\myparagraph{Challenge III: Scarcity of Critical Vulnerability Examples}
Many critical CWEs, particularly among the MITRE Top 25 Most Dangerous CWEs \cite{mitre2024cwe}, are underrepresented in existing datasets. This severe imbalance limits the effectiveness of trained models in identifying rare but high-risk vulnerabilities.

\myparagraph{Summary of Solutions}
To address these challenges, this paper introduces several solutions, including \benchmark for benchmarking and \dataset for training vulnerability detection models.

\myparagraph{Our Solution: \benchmark}
To address \emph{Challenge I}, we introduce \benchmark, a manually curated benchmark focused on the MITRE Top 25 Most Dangerous CWEs \cite{mitre2024cwe} (hereafter, ``Top 25 CWEs''). 
To construct it, we aggregated seven publicly available datasets, performed intra- and inter-dataset deduplication, and standardized CWE annotations based on updated records from the National Vulnerability Database (NVD). 
Due to the large volume of initial data, we applied an LLM-based filtering step to remove non-security-related code changes.
To ensure sufficient coverage, we aimed to curate a balanced benchmark with exactly 50 verified vulnerable samples per CWE category for the Top 25 CWEs. This is important because underrepresented vulnerabilities do not indicate lower danger levels, such as Hard-Coded Credentials (CWE-798), which appear infrequently (see \Cref{fig:cwe25}) but can have catastrophic consequences. In cases where real-world data was insufficient, we introduced the Realistic Vulnerability Generation (\rvg) pipeline, which addresses \emph{Challenge III}. The RVG pipeline utilizes a novel multi-agent LLM workflow that simulates realistic development and security audit processes: (1) \emph{Context \& Threat Modeler} designs practical attack scenarios for the specific CWE; (2) \emph{Vulnerable Implementer} creates corresponding self-contained vulnerable code; (3) \emph{Security Auditor} identifies and remediates the vulnerability; and (4) \emph{Security Reviewer} independently validates both the presence and correct remediation of the target CWE. This process ensures that the synthesized samples are both realistic and targeted to specific CWEs. 
To ensure data quality, we conducted a manual analysis of all candidate samples. 
We hired seven researchers to evaluate each sample against the following criteria: (1) it represents a genuine vulnerability, (2) it is self-contained at the function level, and (3) it is correctly labeled with the intended CWE. We refer to the proportion of samples meeting all three criteria as the benchmark's \emph{correctness}. This validation process resulted in a \emph{correctness} rate of 92\%, yielding a high-quality, balanced, and self-contained benchmark covering the Top 25 CWEs.

\myparagraph{Our Solution: \dataset}
While \benchmark provides a high-quality and reliable evaluation resource, its size is insufficient for training robust machine learning models, motivating the need for a larger and higher-quality training dataset. To solve \emph{Challenge II}, we merged seven publicly available vulnerability datasets and conducted extensive deduplication. To ensure data quality at scale, we applied a novel multi-agent LLM-based pipeline that automatically validated each vulnerability–fix pair. This validation ensured the code change was part of a genuine security fix (in line with its commit message and CVE/CWE description) and not unrelated noise (e.g., refactoring or simple bugs). Specifically, this pipeline consists of independent agents acting as \emph{Auditor}, \emph{Critic}, and \emph{Consensus}: the \emph{Auditor} reviews the evidence for each fix, the \emph{Critic} challenges and verifies the auditor's assessment, and the \emph{Consensus} agent synthesizes these judgments to filter out noisy samples. To prevent data leakage and ensure evaluation integrity, we further removed any overlapping samples between \benchmark and \dataset. Through this multi-stage process, the initial set of 304,726 vulnerability–fix pairs was reduced to a final dataset of 38,548 validated vulnerable functions, which a subsequent manual audit confirmed as 94\% \emph{validity rate} (i.e., the percentage of pairs correctly representing a vulnerability and its corresponding fix) with corresponding fixes suitable for training vulnerability detection models.

\myparagraph{Evaluation}
To assess model generalization, we trained state-of-the-art models on a range of public datasets (including \dataset) and evaluated their performance on our independent, manually verified \benchmark. We distinguish between In-Distribution (ID) performance (testing on the same dataset) and Out-of-Distribution (OOD) performance (testing on \benchmark). Our evaluation further splits \benchmark into its ``Real'' (real-world) and ``Synth'' (synthesized) portions, which we test independently.
Our results reveal a substantial generalization gap (the difference between ID and OOD scores). We find that ID performance does not reliably indicate OOD performance: poor ID could lead to good OOD, while high ID could lead to poor OOD. For example, the Qwen2.5-Coder-1.5B model trained on BigVul achieves a high 0.703(11) ID accuracy but fails on \benchmark's ``Real'' and ``Synth'' portion with 0.493(11) and 0.524(9) accuracy, indicating overfitting.
In contrast, the same model trained on our \dataset achieves a modest 0.590(3) ID score but the highest OOD performance on both the ``Real'' (0.881(26)) and ``Synth'' (0.785(7)) portions of \benchmark. Moreover, by using \rvg to mitigate the CWE imbalance issue (\emph{Challenge III}), augmenting \dataset (RQ3) further enhances generalization, improving performance on real-world data by 5.8\% (from 0.881 to 0.932).

\myparagraph{Main Contributions}
Our main contributions are:

\begin{itemize}[leftmargin=10pt, itemindent=0em]
    \item \textbf{\benchmark}, the first manually verified benchmark covering the Top 25 Most Dangerous CWEs, with 50 vulnerable functions and their fixes per weakness, yielding a total of over 1,000 verified vulnerable functions.

    \item \textbf{\dataset}, a large-scale (38,548 functions), high-quality training dataset curated from seven public sources using rigorous deduplication and a novel multi-agent LLM verification pipeline to ensure high quality.

    \item Realistic Vulnerability Generation (\rvg) pipeline, a novel multi-agent approach to synthesize realistic, context-aware data for underrepresented CWEs. We used \rvg to generate synthetic and realistic samples for \benchmark, specifically targeting the Top 25 CWEs with insufficient representation.

    \item A large-scale empirical study that quantifies intra- and inter-dataset duplication and CWE coverage across major vulnerability datasets, revealing fundamental issues in current resources.
\end{itemize}

This paper is organized as follows: Section~\ref{sec:empirical_study} presents an analysis of existing vulnerability datasets. Section~\ref{sec:benchmark} and Section~\ref{sec:titanvul} detail \benchmark and \dataset. Section~\ref{sec:setup} and Section~\ref{sec:results} show experimental setup and results. Section~\ref{sec:discussion} discusses key implications. Section~\ref{sec:related_work} reviews related literature, and Section~\ref{sec:conclusion} concludes the paper.

\input{figures/top_cwe}

\vspace{-1mm}
\section{Empirical Study}
\label{sec:empirical_study}

We first study the characteristics of seven publicly available function-level vulnerability datasets: BigVul \cite{fan2020ac}, CleanVul \cite{li2024cleanvul}, CVEfixes \cite{bhandari2021cvefixes}, DiverseVul \cite{chen2023diversevul}, PrimeVul \cite{ding2024vulnerability}, SafeCoder \cite{he2024instruction}, and VulnPatchPairs \cite{risse2024uncovering}. We specifically focus on the intra-dataset duplications, inter-dataset duplications, and distributions of CWE types.
This analysis helps us understand the limitations of the publicly available vulnerability datasets and provides the foundation for our benchmark and dataset construction.

\begin{table}[b]
\centering
\vspace{-2mm}
\caption{Vulnerability deduplication analysis.}
\vspace{-2mm}
\label{tb:duplication}
\resizebox{\columnwidth}{!}{%
\begin{tabular}{l|ccc|ccc}
\toprule
\multirow{2}{*}{\textbf{Dataset}} & 
\multicolumn{3}{c|}{\textbf{Complete Pair Duplication}} & 
\multicolumn{3}{c}{\textbf{Self-Identical Duplication}} \\
\cmidrule(lr){2-4}
\cmidrule(lr){5-7}
& \textbf{Initial} & \textbf{After} & \textbf{Removed (\%)} & 
\textbf{Remain} & \textbf{After} & \textbf{Removed (\%)} \\
\midrule
\textbf{BigVul} & 
188,635 & 188,472 & 163 (0.09\%) & 
188,472 & 10,563 & 177,909 (94.40\%) \\
\textbf{CleanVul} & 
42,063 & 42,036 & 27 (0.06\%) & 
42,036 & 41,400 & 636 (1.51\%) \\
\textbf{CVEfixes} & 
41,829 & 19,246 & 22,583 (53.99\%) & 
19,246 & 17,655 & 1,591 (8.27\%) \\
\textbf{DiverseVul} & 
14,484 & 14,471 & 13 (0.06\%) & 
14,471 & 13,805 & 666 (4.60\%) \\
\textbf{PrimeVul} & 
4,704 & 4,700 & 4 (0.09\%) & 
4,700 & 4,698 & 2 (0.04\%) \\
\textbf{SafeCoder} & 
1,268 & 1,251 & 17 (1.34\%) & 
1,251 & 1,238 & 13 (1.04\%) \\
\textbf{VulnPatchPairs} & 
11,743 & 11,743 & 0 (0.00\%) & 
11,743 & 11,377 & 366 (3.12\%) \\
\midrule
\textbf{Total} & 
304,726 & 281,919 & 22,807 (7.48\%) & 
281,919 & 100,736 & 181,183 (64.28\%) \\
\bottomrule
\end{tabular}
}
\vspace{-3mm}
\end{table}

\vspace{-1mm}
\subsection{Intra-Dataset Duplications}
\label{sec:intra_dataset}

Duplication presents an important challenge in vulnerability datasets, potentially skewing analysis results and model performance \citep{ding2024vulnerability}. We examined duplication rates across the seven datasets, identifying two types: 1) Complete Pair Duplication (entire vulnerable-fixed code pairs appearing multiple times) and 2) Self-Identical Duplication (vulnerable code identical to its fixed version). For duplication detection, we used an Abstract Syntax Tree (AST) based approach. We parsed code into AST representations, normalized them by removing comments, and then compared the resulting tree structures.
When a complete pair was found to be duplicated, we retained the pair with more complete metadata (e.g., commit messages, CWE information) and removed the others. This process ensures we resolve duplications while maximizing the information content of the dataset. In cases of self-identical duplication, the pair was removed entirely since it provides no learning signal. The results of these main deduplication stages are presented in \Cref{tb:duplication}.

\myparagraph{Widespread Duplication Across Datasets}
Our analysis reveals that duplication is a widespread issue across publicly available vulnerability datasets. In total, 22,807 redundant pairs (7.48\%) were removed, with the highest rate of complete pair duplication observed in CVEfixes (53.99\%). The second filtering stage eliminated 181,183 self-identical pairs (64.28\% of the remaining corpus), primarily from BigVul (94.40\% of its pairs). After these two rounds of duplication removal, the number of vulnerable–fixed pairs was reduced from 304,726 to 100,736, a total reduction of 66.94\%. These results indicate that intra-dataset duplication is widespread, underscoring the need for rigorous data validation.

\vspace{-1mm}
\subsection{Inter-Dataset Duplications}
\label{sec:cross_dataset}

In addition to intra-dataset duplication, inter-dataset duplication can impact the validity and uniqueness of datasets. Overlapping samples between different datasets can artificially inflate evaluation results and reduce the generalizability of vulnerability detection models. As shown in \Cref{fig:duplication_matrix_heatmap}, duplication rates vary notably: 71.1\% of samples from PrimeVul are present in DiverseVul, while PrimeVul overlaps with CVEfixes (28.4\%). By contrast, many other pairs (e.g., those involving VulnPatchPairs) have less than 1\% overlap, indicating that some datasets remain largely distinct.

\input{figures/heatmap}

\vspace{-1mm}
\subsection{Distribution of CWE Types}

A dataset's CWE distribution is critical; a skewed dataset limits a model's ability to generalize. Understanding this distribution is essential for interpreting model performance and designing fair benchmarks. Figure~\ref{fig:cwe_distribution} presents the distribution of labeled CWE types across six major vulnerability datasets. VulnPatchPairs is not included because it does not provide CWE information. Each subplot displays the frequency of each CWE, sorted in descending order. CWEs classified among the MITRE Top 25 \cite{mitre2024cwe} are highlighted in \colorbox{color-b}{\raisebox{-0.35ex}{\footnotesize dark green}} (others are in \colorbox{color-e}{\raisebox{-0.35ex}{\footnotesize orange}}).

\myparagraph{Significant Imbalances and Dataset-Specific Biases}
In \Cref{fig:cwe_distribution}, the top 5-10 CWEs often comprise 55-80\% of all samples. The frequency ratios between the most and least common CWEs are severe, ranging from approximately 17:1 (DiverseVul) to 192:1 (CleanVul), which can bias model training. Each dataset also shows clear biases: BigVul and PrimeVul are dominated by memory vulnerabilities (CWE-119, 125, 787), while CleanVul, CVEfixes, and SafeCoder are heavily skewed toward web and SQL injection vulnerabilities (CWE-79, 89). These biases mean models trained on one dataset may fail to generalize to vulnerabilities prevalent in others.

\begin{figure}[t]
\vspace{-2mm}
\centering
\scalebox{0.6}{
\begin{tikzpicture}
    \centering
    \begin{axis}[
        height=6.5cm, width=12.5cm,
        /pgf/bar width=0.26cm,
        xmin=1.2, xmax=22.8,
        axis x line*=bottom, axis y line*=left, enlarge x limits=true,
        xtick={0,1,2,3,4,5,6,7,8,9,10,11,12,13,14,15,16,17,18,19,20,21,22,23,24},
        xticklabel style={yshift=-0.8mm, font=\scriptsize, align=center, rotate=45, anchor=east},
        ybar=3.8pt, bar shift=0pt, clip=false,
        ymin=0, ymax=6000,
        ytick={0, 1500, 3000, 4500, 6000},
        yticklabels={0, 1500, 3000, 4500, 6000},
        ymajorgrids, major grid style={draw=black!20}, tick align=inside,
        yticklabel style={font=\footnotesize}, tickwidth=0pt,
        y axis line style={opacity=0},
        ylabel={\normalsize Number of CWE Types},
        y label style={at={(0.21, 1.1)}, rotate=-90},
        xticklabels={
            20, 119, 79, 787, 125, 
            89, 200, 476, 416, 190, 
            400, 94, 918, 22, 352, 
            78, 287, 863, 269, 862, 
            502, 77, 434, 306, 798
        },
    ]
    
    \addplot [draw=light-gray, line width=0.7pt, fill=color-b, error bars/.cd, y dir=both, y explicit, error bar style={draw=black}] coordinates {
        (0, 5063)
        (1, 3777)
        (2, 3711)
        (3, 3651)
        (4, 3379)
        (5, 2589)
        (6, 1792)
        (7, 1785)
        (8, 1755)
        (9, 1373)
        (10, 999)
        (11, 946)
        (12, 764)
        (13, 713)
        (14, 564)
        (15, 451)
        (16, 390)
        (17, 343)
        (18, 280)
        (19, 198)
        (20, 196)
        (21, 134)
        (22, 114)
        (23, 100)
        (24, 39)
    };
    
    \node[above, font=\scriptsize] at (axis cs:0, 5063) {5063};
    \node[above, font=\scriptsize] at (axis cs:1, 3777) {3777};
    \node[above, font=\scriptsize] at (axis cs:2, 3711) {3711};
    \node[above, font=\scriptsize] at (axis cs:3, 3651) {3651};
    \node[above, font=\scriptsize] at (axis cs:4, 3379) {3379};
    \node[above, font=\scriptsize] at (axis cs:5, 2589) {2589};
    \node[above, font=\scriptsize] at (axis cs:6, 1792) {1792};
    \node[above, font=\scriptsize] at (axis cs:7, 1785) {1785};
    \node[above, font=\scriptsize] at (axis cs:8, 1755) {1755};
    \node[above, font=\scriptsize] at (axis cs:9, 1373) {1373};
    \node[above, font=\scriptsize] at (axis cs:10, 999) {999};
    \node[above, font=\scriptsize] at (axis cs:11, 946) {946};
    \node[above, font=\scriptsize] at (axis cs:12, 764) {764};
    \node[above, font=\scriptsize] at (axis cs:13, 713) {713};
    \node[above, font=\scriptsize] at (axis cs:14, 564) {564};
    \node[above, font=\scriptsize] at (axis cs:15, 451) {451};
    \node[above, font=\scriptsize] at (axis cs:16, 390) {390};
    \node[above, font=\scriptsize] at (axis cs:17, 343) {343};
    \node[above, font=\scriptsize] at (axis cs:18, 280) {280};
    \node[above, font=\scriptsize] at (axis cs:19, 198) {198};
    \node[above, font=\scriptsize] at (axis cs:20, 196) {196};
    \node[above, font=\scriptsize] at (axis cs:21, 134) {134};
    \node[above, font=\scriptsize] at (axis cs:22, 114) {114};
    \node[above, font=\scriptsize] at (axis cs:23, 100) {100};
    \node[above, font=\scriptsize] at (axis cs:24, 39) {39};
    \end{axis}
\end{tikzpicture}
}
\vspace{-1mm}
\caption{Distribution of MITRE top 25 most dangerous CWE across the consolidated vulnerability dataset.}
\label{fig:cwe25}
\vspace{-3.5mm}
\end{figure}

\myparagraph{Challenges of Severe CWE Imbalance}
We merged all datasets to examine the consolidated distribution, shown in \Cref{fig:cwe25}. This consolidated view confirms a severe imbalance. The frequency ratio between the most common type, CWE-20 (5,063 samples), and the least common, CWE-798 (39 samples), is approximately 130:1. This skew underscores the challenge for models to generalize and shows the difficulty of using these datasets directly as balanced benchmarks, as many dangerous CWEs are severely underrepresented.

\vspace{-1mm}
\section{\benchmark: A Benchmark for the Top 25 Most Dangerous CWE Weaknesses}
\label{sec:benchmark}

The construction of \benchmark, a comprehensive benchmark for evaluating vulnerability detection approaches across the MITRE Top 25 Most Dangerous CWEs \cite{mitre2024cwe}, follows a multi-stage approach, as illustrated in \Cref{fig:benchvul}. We detail each stage of this process below.

\vspace{-1mm}
\subsection{Data Integration}

We first aggregated multiple publicly available vulnerability datasets, including BigVul \cite{fan2020ac}, CleanVul \cite{li2024cleanvul}, CVEfixes \cite{bhandari2021cvefixes}, DiverseVul \cite{chen2023diversevul}, PrimeVul \cite{ding2024vulnerability}, SafeCoder \cite{he2024instruction}, and VulnPatchPairs \cite{risse2024uncovering}. We then standardized these datasets into a unified format. Manual inspection revealed inconsistencies in CWE labeling compared to the NVD, largely due to outdated or missing annotations. We thus updated each vulnerability's CWE information by retrieving the latest annotations from the NVD based on CVE identifiers, using data as of December 5, 2024. We observed that initially, only 74.83\% of PrimeVul, 43.01\% of DiverseVul, and 70.97\% of BigVul samples matched the NVD's CWE annotations. Across all datasets, 12,127 vulnerabilities had CWE labels updated. Additionally, datasets lacking CWE annotations, such as CVEfixes, were supplemented by deriving CWE identifiers directly from their corresponding CVE records. In cases where a CVE was associated with multiple CWEs in the NVD records, we retained all associated CWE labels for that vulnerability. Furthermore, we analyzed the MITRE Top 25 CWEs for hierarchical or logical conflicts, refining the set to 21 distinct types, and the full details are available in our replication package. We then applied intra-dataset deduplication (\Cref{sec:intra_dataset}), reducing the set from 304,726 to 100,736 pairs. Next, we merged the cleaned datasets and performed inter-dataset deduplication (\Cref{sec:cross_dataset}) to obtain a unified vulnerability dataset.

\vspace{-1mm}
\subsection{LLM-Based Filtering}

To construct a high-quality, function-level benchmark covering the MITRE Top 25 Most Dangerous CWEs~\cite{mitre2024cwe}, each vulnerability-fixing pair should be manually verified. Specifically, we aim to ensure that each pair accurately represents a genuine vulnerability fix and is self-contained, meaning the vulnerability fix can be fully understood by examining only the code within a single function~\cite{risse2025top}. However, manually validating every pair is impractical due to the large volume of samples available for some CWEs (e.g., over 5,000 instances as shown in \Cref{fig:cwe25}). Furthermore, prior studies indicate that a significant portion of labeled vulnerabilities do not genuinely address security flaws but rather represent unrelated bug fixes, refactoring, or other code changes~\cite{ding2024vulnerability,chen2023diversevul}. 
To efficiently address these challenges, we first leverage LLMs to filter out unrelated code changes, substantially reducing the number of candidate samples requiring manual verification~\cite{li2024cleanvul}. Although LLM-based filtering can occasionally introduce false positives or negatives, we mitigate this risk through subsequent structured manual reviews (see Section~\ref{sec:manual_review}), where each remaining sample is carefully checked to confirm it represents a genuine and self-contained vulnerability fix. This combined approach ensures the final benchmark maintains high accuracy while significantly improving validation efficiency.

\begin{figure*}[t]
\centering
\includegraphics[width=\linewidth]{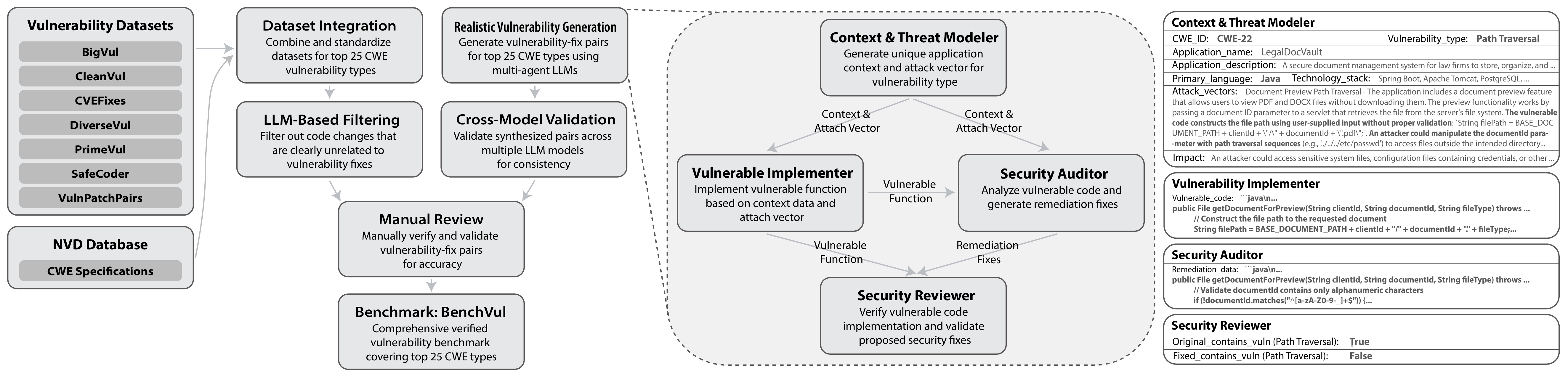}
\vspace{-3mm}
\caption{Overview of the \benchmark construction pipeline for the MITRE Top 25 Most Dangerous CWEs.}
\label{fig:benchvul}
\vspace{-3mm}
\end{figure*}

\vspace{-1mm}
\subsection{Realistic Vulnerability Generation}
\label{sec:rvg}

Constructing a robust benchmark of 50 vulnerability-fix pairs per CWE requires sufficient self-contained, real-world examples. However, since this data is lacking for some CWE types, we synthesize realistic pairs to fill the gaps. To address this, we propose the \textbf{Realistic Vulnerability Generation} (\rvg) pipeline, a novel multi-agent LLM approach illustrated in the right of \Cref{fig:benchvul}. \rvg comprises four interrelated roles, detailed as follows.
\emph{Context \& Threat Modeler}, \emph{Vulnerable Implementer}, \emph{Security Auditor}, and \emph{Security Reviewer}. Each role contributes to generating realistic, validated vulnerability pairs, detailed as follows. 
Due to space constraints, the full prompts and scripts are provided in the replication package\footnote{\url{https://github.com/yikun-li/TitanVul-BenchVul}}.

\myparagraph{Context \& Threat Modeler} 
Given a CWE ID and description as inputs, this agent initiates the \rvg process by creating a realistic application context and identifying a corresponding attack vector. To maximize diversity and realism, this agent selects a distinct programming language, technology stack, user roles, and functionalities for each scenario. It also maintains uniqueness by tracking previously generated contexts, employing a first-in-first-out (FIFO) approach to prevent repetition.

\myparagraph{Vulnerable Implementer} 
This agent generates a realistic and self-contained vulnerable code snippet based explicitly on the context and attack vector defined by the previous agent. The code incorporates subtle but exploitable vulnerabilities, accompanied by comments describing the intended functionality without indicating vulnerabilities. 

\myparagraph{Security Auditor} 
The Security Auditor analyzes the context, attack vector, and vulnerable code snippet to identify security flaws and subsequently produces a secure version of the code. 

\myparagraph{Security Reviewer} 
This agent performs a comparative evaluation of the vulnerable and remediated code snippets. It verifies whether the identified CWE-related vulnerability is present in the vulnerable snippet and mitigated.

\vspace{-1mm}
\subsection{Cross-Model Validation}

To strengthen the robustness of synthesized vulnerability data, we conducted cross-model validation using different state-of-the-art LLMs. Specifically, we utilized Claude-3.7-Sonnet for initial synthesis tasks and GPT-4o for validation purposes. Each synthesized vulnerability-fixing pair generated by Claude-3.7-Sonnet was independently assessed by GPT-4o, verifying whether the vulnerability was correctly implemented and remediated.

\vspace{-1mm}
\subsection{Manual Review}
\label{sec:manual_review}

After the automated filtering and synthesis stages, all vulnerability-fix pairs underwent human review. 
This initial human review was conducted by a single annotator, who examined all pairs to manually verify that every pair (1) represents a genuine vulnerability, (2) is self-contained at the function level (i.e., can be understood without external context), and (3) is correctly labeled with the intended CWE.
As most previous research in this area has focused on function-level vulnerability detection \cite{risse2023limits,risse2025top}, we aimed to build a high-quality, self-contained, balanced, independent benchmark for evaluating the generalization capabilities of these models.
This function-level, self-contained criterion is critical for building an independent benchmark; however, it also presents a challenge.
As noted in previous work \cite{risse2025top}, many real-world vulnerabilities are inter-procedural, making it difficult to find high-quality, self-contained examples.
Our manual verification process yielded uneven numbers of real-world samples across CWEs. For instance, we were able to identify 50 high-quality samples for CWE-89, 50 for CWE-78, 49 for CWE-79, 18 for CWE-22, 16 for CWE-502, 6 for CWE-94, and 1 for CWE-863.
In total, this process yielded 190 real-world samples.
Our goal was to create a benchmark with a balanced number of samples per CWE, but the scarcity of high-quality real-world data motivated our use of \rvg.
To fill the gaps for underrepresented CWEs and meet our target, synthesized pairs produced by \rvg were used. These synthesized pairs were then reviewed using the same manual criteria to ensure quality.
This procedure yielded a final benchmark of 1,050 vulnerable functions and their 1,050 corresponding remediations, balanced across the top CWEs. In total, 190 of these vulnerable samples (18.1\%) are from real-world data, and the remaining 860 (81.9\%) are synthetic samples produced and validated by \rvg.

To assess benchmark quality, seven independent researchers with experience in vulnerability analysis participated in the review. 
Each researcher was assigned a random sample of the benchmark and asked to evaluate whether the vulnerabilities met the aforementioned criteria. 
Out of 275 reviewed pairs, 253 were correct, for an overall \emph{correctness} rate of 92\%.
We calculated Cohen's Kappa for inter-rater reliability, which was 0.453, indicating moderate agreement.
This value may be partly attributed to the low prevalence of incorrect samples (a known statistical issue called the kappa paradox \cite{wan2015kappa}), as the overall correctness rate was high.
This level of agreement is also consistent with previous software engineering studies \cite{wang2025towards,wei2021comprehensive}.
While the benchmark labels are not perfect, achieving a \emph{correctness} rate above 90\% is generally considered sufficient for reliable evaluation in empirical software engineering research~\cite{ray2014large}, providing confidence that our benchmark supports reliable empirical evaluation.

\input{figures/similarity}

\myparagraph{Semantic Similarity Analysis}
To further validate \benchmark's independence and check for potential data leakage, we conducted a semantic similarity analysis. We adopted a standard embedding-based approach by using the pre-trained UniXcoder model \cite{guo2022UniXcoder} to encode each vulnerable function into a semantic vector embedding. We then computed the average cosine similarity between all pairs of functions from every two datasets \cite{zhang2024enhancing}.
The results are presented in the heatmap in \Cref{fig:smililarity_matrix_heatmap}. This analysis confirms that both the ``Real'' (real-world) and ``Synth'' (synthesized) portions of \benchmark have low average semantic similarity scores (generally between 0.33 and 0.38) when compared to all other training datasets. The similarity between ``Real'' and ``Synth'' themselves is also low (0.3604). This lack of high semantic overlap indicates that \benchmark is a sufficiently independent testbed, and that our synthesized data is not merely a semantic paraphrase of existing data. Interestingly, this semantic analysis also confirms our earlier AST-based duplication findings (from \Cref{fig:duplication_matrix_heatmap}), showing a high similarity score between PrimeVul and DiverseVul (0.4910). This consistency further validates our similarity analysis method.

\vspace{-1mm}
\section{\dataset: A Large-Scale and High-Quality Vulnerability Dataset}
\label{sec:titanvul}

Vulnerability detection models require not only evaluation benchmarks, but also large, high-quality training datasets.
While \benchmark offers manually verified data for evaluation, its limited scale and the cost of manual validation make it impractical for training models. 
Thus, scalable methods are needed to curate high-quality, noise-free vulnerability data for effective model training.
Existing vulnerability datasets vary widely in quality. Prior studies report that only a fraction of samples in several popular datasets represent valid vulnerability fixes, where \emph{validity rate} or \emph{validity} is defined as the percentage of vulnerability-fix code pair associated with vulnerability fixes \cite{ding2024vulnerability,chen2023diversevul}: \textit{BigVul} (25.0\%), \textit{VulnPatchPairs} (36.0\%), \textit{CVEfixes} (51.7\%), and \textit{DiverseVul} (60.0\%). In contrast, \textit{CleanVul} and \textit{PrimeVul} achieve higher \emph{validity rates} of 90.6\% and 86.0\%, respectively. 

This low \emph{validity} is largely because many of these datasets contain significant noise from samples that are unrelated to security, such as test code, code refactoring, or simple bug fixes that were part of the same commit \cite{ding2024vulnerability}.
For a sample to be a truly effective training example, it should represent a genuine security fix that is clearly understandable from the CWE label, commit message, and code diff. If this context is missing or incorrect, such samples are unlikely to help models learn true vulnerability detection.
While \benchmark provides a high-quality, manually verified set of over 1,000 self-contained vulnerability–fix pairs, its limited size and the resource-intensive nature of manual validation make it impractical to use as a large-scale training dataset.
To develop robust and generalizable vulnerability detection models, there is a clear need for larger, high-quality datasets that are both reliable and scalable.

\begin{figure}[t]
\centering
\includegraphics[width=0.81\linewidth]{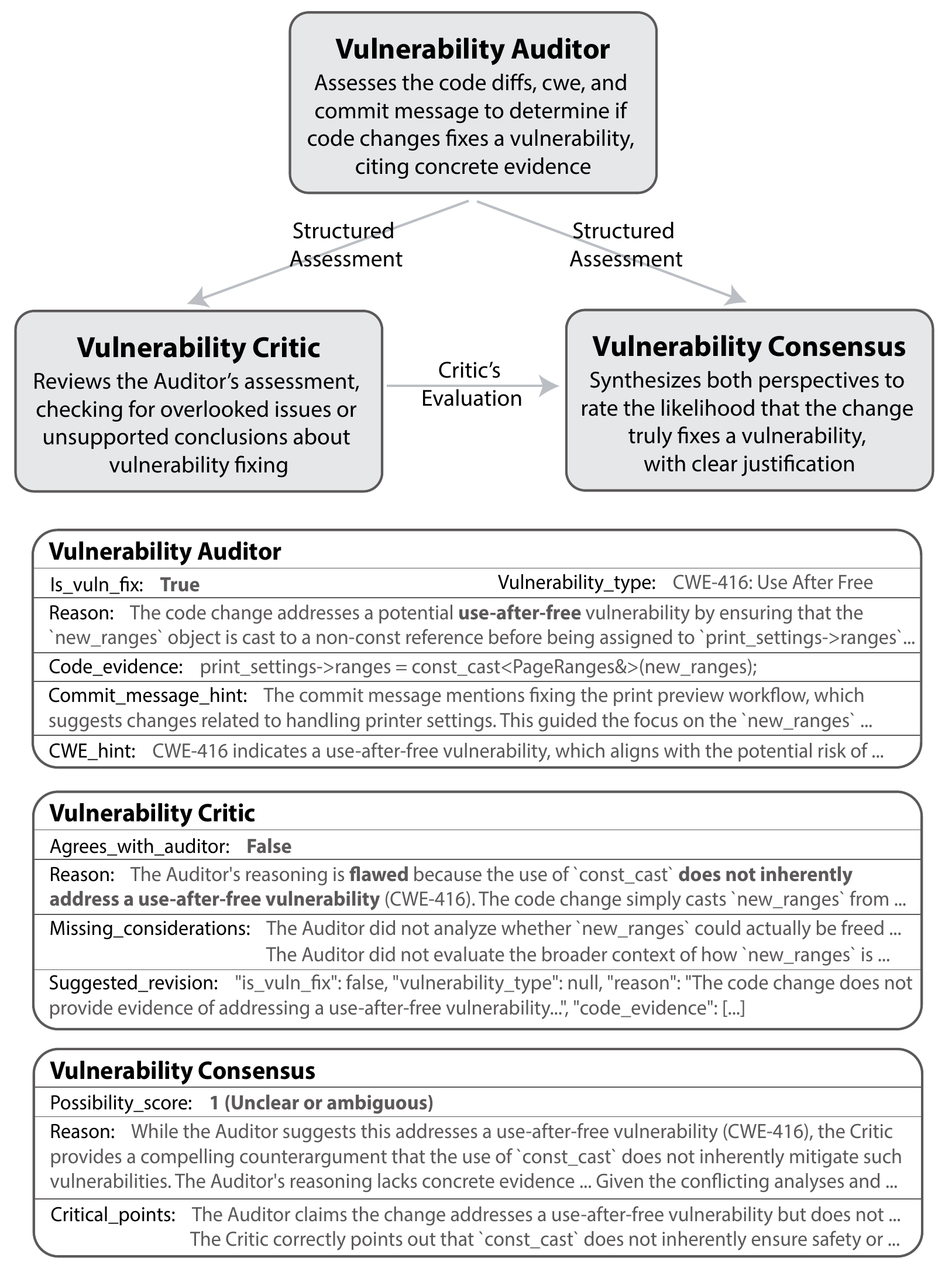}
\vspace{-1mm}
\caption{Overview of the multi-agent LLM verification pipeline used to construct \dataset.}
\label{f:framwork_data}
\vspace{-5mm}
\end{figure}

To address this, we re-examined the consolidated dataset (in Section \ref{sec:empirical_study}) with the primary goal of removing non-security-related noise.
We automated this process using a novel multi-agent pipeline leveraging LLMs for analysis, verification, and validation of security vulnerabilities. The architecture of the pipeline is illustrated in \Cref{f:framwork_data}, comprising three key components: \emph{Vulnerability Auditor}, \emph{Vulnerability Critic}, and \emph{Vulnerability Consensus}.

\myparagraph{Vulnerability Auditor} This agent serves as the initial evaluator, analyzing code diffs, commit messages, and associated CWE information. Its primary role is to determine whether the code changes represent genuine security vulnerability fixes. The Auditor provides detailed evidence by identifying the type of vulnerability addressed, highlighting relevant code snippets, and incorporating insights from commit messages or CWE hints. 

\myparagraph{Vulnerability Critic} The agent conducts a second review, carefully examining the Auditor's findings to ensure their accuracy, completeness, and reliability. It identifies any overlooked issues, incorrect reasoning, or weak evidence in the Auditor's analysis. By providing constructive feedback and corrections, the Critic ensures a thorough and reliable evaluation of each vulnerability fix.

\myparagraph{Vulnerability Consensus} This agent synthesizes the analyses from the Auditor and Critic to produce a unified and justified assessment. It assigns a possibility score (ranging from 0 to 3) indicating the likelihood that the code change genuinely addresses a security vulnerability. This consensus-building process carefully considers both agreement and disagreement points among previous analyses, prioritizing concrete evidence and clearly articulating its reasoning.

\begin{figure}[b]
\vspace{-2mm}
\newcommand{\semitimy}{%
  \fontsize{6pt}{7pt}\selectfont
}
\centering
\scalebox{0.6}{
\begin{tikzpicture}
    \centering
    \begin{axis}[
        height=6.5cm, width=12.5cm,
        /pgf/bar width=0.26cm,
        xmin=1.2, xmax=22.8,
        axis x line*=bottom, axis y line*=left, enlarge x limits=true,
        xtick={0,1,2,3,4,5,6,7,8,9,10,11,12,13,14,15,16,17,18,19,20,21,22,23,24},
        xticklabel style={yshift=-0.8mm, font=\scriptsize, align=center, rotate=45, anchor=east},
        ybar=3.8pt, bar shift=0pt, clip=false,
        ymin=0, ymax=1800,
        ytick={0, 300, 600, 900, 1200, 1500, 1800},
        yticklabels={0, 300, 600, 900, 1200, 1500, 1800},
        ymajorgrids, major grid style={draw=black!20}, tick align=inside,
        yticklabel style={font=\footnotesize}, tickwidth=0pt,
        y axis line style={opacity=0},
        ylabel={\normalsize Number of CWE Types},
        y label style={at={(0.21, 1.1)}, rotate=-90},
        xticklabels={
            787, 20, 119, 125, 79, 
            200, 416, 476, 190, 94, 
            22, 89, 400, 918, 863, 
            352, 78, 269, 287, 502, 
            862, 77, 434, 798, 306
        },
    ]
    
    \addplot [draw=light-gray, line width=0.7pt, fill=color-b, error bars/.cd, y dir=both, y explicit, error bar style={draw=black}] coordinates {
        (0, 1846)
        (1, 1734)
        (2, 1520)
        (3, 1432)
        (4, 968)
        (5, 789)
        (6, 767)
        (7, 721)
        (8, 570)
        (9, 540)
        (10, 327)
        (11, 325)
        (12, 296)
        (13, 218)
        (14, 176)
        (15, 171)
        (16, 168)
        (17, 144)
        (18, 138)
        (19, 68)
        (20, 63)
        (21, 56)
        (22, 46)
        (23, 16)
    };
    
    \node[above, font=\semitimy] at (axis cs:0, 1846) {1846};
    \node[above, font=\semitimy] at (axis cs:1, 1734) {1734};
    \node[above, font=\semitimy] at (axis cs:2, 1520) {1520};
    \node[above, font=\semitimy] at (axis cs:3, 1432) {1432};
    \node[above, font=\semitimy] at (axis cs:4, 968) {968};
    \node[above, font=\semitimy] at (axis cs:5, 789) {789};
    \node[above, font=\semitimy] at (axis cs:6, 767) {767};
    \node[above, font=\semitimy] at (axis cs:7, 721) {721};
    \node[above, font=\semitimy] at (axis cs:8, 570) {570};
    \node[above, font=\semitimy] at (axis cs:9, 540) {540};
    \node[above, font=\semitimy] at (axis cs:10, 327) {327};
    \node[above, font=\semitimy] at (axis cs:11, 325) {325};
    \node[above, font=\semitimy] at (axis cs:12, 296) {296};
    \node[above, font=\semitimy] at (axis cs:13, 218) {218};
    \node[above, font=\semitimy] at (axis cs:14, 176) {176};
    \node[above, font=\semitimy] at (axis cs:15, 171) {171};
    \node[above, font=\semitimy] at (axis cs:16, 168) {168};
    \node[above, font=\semitimy] at (axis cs:17, 144) {144};
    \node[above, font=\semitimy] at (axis cs:18, 138) {138};
    \node[above, font=\semitimy] at (axis cs:19, 68) {68};
    \node[above, font=\semitimy] at (axis cs:20, 63) {63};
    \node[above, font=\semitimy] at (axis cs:21, 56) {56};
    \node[above, font=\semitimy] at (axis cs:22, 46) {46};
    \node[above, font=\semitimy] at (axis cs:23, 16) {16};
    \node[above, font=\semitimy] at (axis cs:24, 5) {5};
    \end{axis}
\end{tikzpicture}
}
\vspace{-3mm}
\caption{Distribution of MITRE top 25 most dangerous CWE across \dataset.}
\label{fig:cwe25_titanvul}
\vspace{-5mm}
\end{figure}

\myparagraph{\dataset}
We begin by performing comprehensive deduplication and merging of datasets and updating CWE labels. Next, we employ our multi-agent pipeline to further enhance data quality. To validate the quality of \dataset, six researchers manually audited 400 randomly selected vulnerability-fix pairs to check for noise. We define \emph{validity rate} or \emph{validity} as the percentage of pairs representing a genuine vulnerability consistent with its CVE description or commit message. This audit confirmed a \emph{validity rate} of 94\%. For inter-rater reliability, we calculated Cohen's Kappa of 0.424, indicating moderate agreement \cite{wan2015kappa}. This value may be partly attributed to the low prevalence of incorrect samples (a known statistical issue called the ``kappa paradox'' \cite{cicchetti1990high}), as the overall \emph{validity rate} was high. Similar agreement levels have been reported in previous software engineering studies \cite{wang2025towards,wei2021comprehensive}. Finally, to prevent any potential data leakage, we remove duplicate samples between \benchmark and other sources in the final dataset. 
The resulting dataset comprises 38,548 vulnerable functions along with their corresponding fixes, establishing \dataset as a reliable resource for vulnerability-related research.
\Cref{fig:cwe25_titanvul} illustrates the distribution of the MITRE Top 25 CWEs within this final dataset. As the figure shows, the dataset exhibits substantial class imbalance; we mitigate the impact of this imbalance on evaluation by using our balanced \benchmark benchmark (RQ1/2) and on training via \rvg augmentation (RQ3).

\vspace{-1mm}
\section{Experimental Setup}
\label{sec:setup}
\begin{table*}[!htbp]
\centering
\caption{Performance of language models evaluated on \benchmark and on their respective source datasets.}
\vspace{-2mm}
\label{tb:model_performance}
\resizebox{\textwidth}{!}{%
\begin{tabular}{
l|
p{1.3cm}<{\centering}p{1.3cm}<{\centering}p{1.3cm}<{\centering}|
p{1.3cm}<{\centering}p{1.3cm}<{\centering}p{1.3cm}<{\centering}|
p{1.3cm}<{\centering}p{1.3cm}<{\centering}p{1.3cm}<{\centering}|
p{1.3cm}<{\centering}p{1.3cm}<{\centering}p{1.3cm}<{\centering}
}
\toprule
\multirow{2}{*}{\textbf{Model}} & 
\multicolumn{3}{c|}{\textbf{Trained on BigVul}} & 
\multicolumn{3}{c|}{\textbf{Trained on CVEfixes}} & 
\multicolumn{3}{c|}{\textbf{Trained on CleanVul}} & 
\multicolumn{3}{c}{\textbf{Trained on DiverseVul}} \\
\cmidrule(lr){2-4} \cmidrule(lr){5-7} \cmidrule(lr){8-10} \cmidrule(lr){11-13}
& \textbf{ID} & \textbf{Real} & \textbf{Synth} &
\textbf{ID} & \textbf{Real} & \textbf{Synth} &
\textbf{ID} & \textbf{Real} & \textbf{Synth} &
\textbf{ID} & \textbf{Real} & \textbf{Synth} \\
\midrule

\textbf{CodeBERT} &
\num{0.615(5)} & \num{0.501(5)} & \num{0.522(5)} &
\num{0.509(4)} & \cellcolor{color-f}\num{0.700(6)} & \num{0.606(24)} &
\num{0.534(22)} & \num{0.641(97)} & \num{0.650(124)} &
\num{0.500(0)} & \num{0.500(0)} & \num{0.500(0)} \\

\textbf{GraphCodeBERT} & 
\num{0.615(3)} & \num{0.506(7)} & \num{0.520(3)} &
\num{0.509(7)} & \cellcolor{color-f}\num{0.745(17)} & \num{0.613(24)} &
\num{0.541(20)} & \num{0.669(63)} & \num{0.634(103)} &
\num{0.507(6)} & \num{0.511(11)} & \num{0.576(118)} \\

\textbf{UniXcoder} & 
\num{0.667(7)} & \textbf{\num{0.519(7)}} & \textbf{\num{0.529(11)}} &
\num{0.512(6)} & \cellcolor{color-f}\num{0.766(25)} & \num{0.660(48)} &
\num{0.566(6)} & \cellcolor{color-f}\num{0.784(18)} & \cellcolor{color-f}\num{0.742(6)} &
\num{0.526(3)} & \num{0.540(14)} & \cellcolor{color-f}\num{0.717(11)} \\

\textbf{Llama-3.2-3B} & 
\num{0.677(8)} & \cellcolor{color-e}\num{0.496(9)} & \num{0.528(14)} &
\num{0.511(6)} & \cellcolor{color-f}\num{0.777(8)} & \num{0.662(38)} &
\textbf{\num{0.571(5)}} & \cellcolor{color-f}\num{0.776(20)} & \num{0.752(9)} &
\num{0.508(1)} & \num{0.520(11)} & \num{0.578(27)} \\

\textbf{Qwen2.5-Coder-1.5B} &
\cellcolor{color-f}\textbf{\num{0.703(11)}} & \cellcolor{color-e}\num{0.493(11)} & \num{0.524(9)} &
\textbf{\num{0.518(2)}} & \cellcolor{color-b}\textbf{\num{0.837(8)}} & \cellcolor{color-f}\textbf{\num{0.702(38)}} &
\num{0.569(9)} & \cellcolor{color-f}\textbf{\num{0.790(21)}} & \cellcolor{color-f}\textbf{\num{0.754(16)}} &
\textbf{\num{0.538(4)}} & \textbf{\num{0.654(3)}} & \cellcolor{color-f}\textbf{\num{0.720(13)}} \\

\midrule
\end{tabular}%
}

\resizebox{\textwidth}{!}{%
\begin{tabular}{
l|
p{1.3cm}<{\centering}p{1.3cm}<{\centering}p{1.3cm}<{\centering}|
p{1.3cm}<{\centering}p{1.3cm}<{\centering}p{1.3cm}<{\centering}|
p{1.3cm}<{\centering}p{1.3cm}<{\centering}p{1.3cm}<{\centering}|
p{1.3cm}<{\centering}p{1.3cm}<{\centering}p{1.3cm}<{\centering}
}

\toprule
\multirow{2}{*}{\textbf{Model}} & 
\multicolumn{3}{c|}{\textbf{Trained on \dataset}} & 
\multicolumn{3}{c|}{\textbf{Trained on PrimeVul}} & 
\multicolumn{3}{c|}{\textbf{Trained on SafeCoder}} & 
\multicolumn{3}{c}{\textbf{Trained on VulnPatchPairs}} \\
\cmidrule(lr){2-4} \cmidrule(lr){5-7} \cmidrule(lr){8-10} \cmidrule(lr){11-13}
& \textbf{ID} & \textbf{Real} & \textbf{Synth} & 
\textbf{ID} & \textbf{Real} & \textbf{Synth} & 
\textbf{ID} & \textbf{Real} & \textbf{Synth} & 
\textbf{ID} & \textbf{Real} & \textbf{Synth} \\
\midrule

\textbf{CodeBERT} & 
\num{0.500(0)} & \num{0.500(0)} & \num{0.500(0)} &
\num{0.518(11)} & \num{0.529(11)} & \num{0.636(113)} &
\num{0.553(22)} & \num{0.593(16)} & \num{0.563(38)} &
\num{0.503(6)} & \num{0.501(2)} & \num{0.504(6)} \\

\textbf{GraphCodeBERT} & 
\num{0.557(2)} & \cellcolor{color-f}\num{0.741(89)} & \cellcolor{color-f}\num{0.712(6)} &
\num{0.526(2)} & \num{0.527(5)} & \num{0.673(20)} &
\num{0.569(7)} & \num{0.618(23)} & \num{0.567(24)} &
\num{0.533(4)} & \num{0.517(17)} & \num{0.546(32)} \\

\textbf{UniXcoder} & 
\num{0.575(4)} & \cellcolor{color-b}\num{0.849(21)} & \cellcolor{color-f}\num{0.749(3)} &
\num{0.538(3)} & \num{0.514(3)} & \cellcolor{color-f}\textbf{\num{0.736(5)}} &
\num{0.574(3)} & \num{0.663(33)} & \num{0.638(28)} &
\num{0.559(2)} & \num{0.507(6)} & \textbf{\num{0.641(18)}} \\

\textbf{Llama-3.2-3B} & 
\num{0.578(4)} & \cellcolor{color-b}\num{0.809(42)} & \cellcolor{color-f}\num{0.766(4)} &
\num{0.508(10)} & \num{0.516(9)} & \num{0.559(56)} &
\textbf{\num{0.595(22)}} & \cellcolor{color-f}\textbf{\num{0.701(87)}} & \textbf{\num{0.643(60)}} &
\num{0.536(7)} & \num{0.512(9)} & \num{0.558(39)} \\

\textbf{Qwen2.5-Coder-1.5B} &
\textbf{\num{0.590(3)}} & \cellcolor{color-b}\textbf{\num{0.881(26)}} & \cellcolor{color-f}\textbf{\num{0.785(7)}} &
\textbf{\num{0.539(6)}} & \textbf{\num{0.545(12)}} & \cellcolor{color-f}\num{0.703(6)} &
\num{0.531(4)} & \num{0.541(36)} & \num{0.546(20)} &
\textbf{\num{0.562(9)}} & \textbf{\num{0.543(15)}} & \num{0.537(24)} \\

\bottomrule
\end{tabular}%
}
\begin{flushleft}
\small
\textbf{Note:} Accuracy is reported (datasets are balanced). Columns: ``ID'' (in-distribution (ID) evaluation: train/test on same dataset), ``Real'' (test on \benchmark's real-world data), ``Synth'' (test on \benchmark's synthetic data). Highest value per column is \textbf{bold}. Highlights: \colorbox{color-b}{\raisebox{-0.35ex}{\footnotesize dark green}} > 0.8, \colorbox{color-f}{\raisebox{0.35ex}{\footnotesize green}} > 0.7, \colorbox{color-e}{\footnotesize orange} < 0.5.
\end{flushleft}
\vspace{-5mm}
\end{table*}

\subsection{Research Questions}

We formulate the following research questions (RQs):

\myparagraph{RQ1: How well can models trained on vulnerability datasets detect the Top 25 Most Dangerous CWEs?}
Due to the lack of dedicated vulnerability benchmarks, most prior studies split a single dataset for training and testing, making it difficult to assess true generalization \cite{risse2025top}. Given the widespread issues of overfitting and dataset bias, it is critical to rigorously evaluate whether models can actually identify the Top 25 CWE weaknesses on an independent and high-quality benchmark (\benchmark).
    
\myparagraph{RQ2: How does the choice of training dataset affect model performance across CWE categories?}
Our analysis reveals that publicly available datasets differ widely in their CWE distribution and quality, with each exhibiting distinct biases toward certain vulnerability types (e.g., memory safety, web security). Understanding how these differences impact model performance can illuminate the strengths and weaknesses of popular datasets and inform future dataset construction and model development.

\myparagraph{RQ3: Does adding synthesized data to the training dataset improve detection of the Top 25 Most Dangerous CWEs?}
Since many most dangerous CWEs are rare in real-world datasets, models may lack sufficient examples to learn robust patterns. Synthetic data generation offers a potential solution by augmenting scarce categories and improving model coverage. Evaluating the actual benefit of synthesized data for detecting critical weaknesses is thus important for advancing practical ML-based vulnerability detection.

\vspace{-1mm}
\subsection{Models}

We evaluate a diverse set of five language models for vulnerability detection, including CodeBERT \cite{feng2020codebert}, GraphCodeBERT \cite{guo2020graphcodebert}, UniXcoder \cite{guo2022UniXcoder}, Llama‑3.2‑3B \cite{grattafiori2024llama}, and Qwen2.5-Coder-1.5B \cite{hui2024qwen2}. 
We set the maximum input token limit to 512 for CodeBERT and GraphCodeBERT, 1,024 for UniXcoder and Llama‑3.2-3B, and 4,096 for Qwen2.5-Coder-1.5B. Inputs exceeding the limits are truncated. 
This selection enables a comprehensive comparison of architectural styles and model scales in the context of vulnerability detection.

\vspace{-1mm}
\subsection{Evaluation Metrics}

We evaluate model performance using two standard metrics: \textbf{accuracy} and \textbf{F1-score}. Accuracy reflects the proportion of correctly classified samples in our balanced dataset. 
Precision and recall measure, respectively, how many predicted vulnerabilities are correct and how many actual vulnerabilities are detected. 
The F1-score, the harmonic mean of precision and recall, provides an overall balance between these two metrics.

\vspace{-1mm}
\subsection{Implementation Details}

We split each dataset into training (70\%), validation (15\%), and test (15\%) sets using a time-aware (temporal) split. This approach ensures that training samples are chronologically older than validation and test samples, simulating a realistic deployment scenario. To implement this split, we obtained date information for each sample. For the CleanVul and CVEfixes datasets, we used the date metadata readily available. For BigVul, DiverseVul, PrimeVul, SafeCoder, and VulnPatchPairs, we cloned the associated repositories and extracted the commit date for each vulnerability. Training is conducted for up to 10 epochs, and the best-performing checkpoints are retained for evaluation. All experiments are run on NVIDIA H100 GPUs with an Intel Xeon Platinum 8480C CPU.
To account for statistical variability, results are averaged over three runs with different random seeds. 
We report the mean and use the concise uncertainty notation (e.g., \num{0.615(5)}) to indicate the standard deviation in the last digit(s).

\vspace{-1mm}
\section{Results}
\label{sec:results}

\subsection{RQ1: Dataset Performance on Top 25 CWEs}

We evaluated the effectiveness of language models trained on various vulnerability datasets in detecting the MITRE Top 25 Most Dangerous CWEs \cite{mitre2024cwe} using our curated benchmark, \benchmark. Table~\ref{tb:model_performance} presents comprehensive results across eight datasets and five model architectures.
We primarily use accuracy as the evaluation metric because \benchmark contains a balanced number of vulnerable and non-vulnerable samples for each dataset, making accuracy straightforward to interpret and directly comparable to the random guessing baseline (0.5). In contrast, metrics such as F1-score can sometimes be misleading. For instance, a naive model that predicts all samples as vulnerable would achieve perfect recall (1.0), precision of 0.5, and thus an inflated F1-score of 0.667, despite performing no better than random guessing.

\myparagraph{The Generalization Gap: In-Distribution (ID) vs. Out-of-Dis\-tri\-bu\-tion (OOD) Performance}
Our analysis in \Cref{tb:model_performance} evaluates models using two distinct setups: In-Distribution (ID) evaluation, which refers to training and testing on splits from the same dataset, and Out-of-Distribution (OOD) evaluation, which involves testing on our independent \benchmark. This OOD evaluation is further divided into \benchmark's ``Real'' (real-world) and ``Synth'' (synthesized) data portions. We observe a clear generalization gap, which we define as the performance difference between these two setups: $Generalization Gap = Performance_{ID} - Performance_{OOD}$. A large, positive gap suggests a model has overfitted to dataset-specific artifacts, whereas a small or negative gap indicates better generalization.

To compare dataset performance consistently, we primarily reference the results from Qwen2.5-Coder-1.5B, as it shows strong performance among all models. 
We notice that the ID performance has almost no clear relation to the OOD performance on the independent \benchmark. This is most evident with BigVul, which achieves the highest ID accuracy (0.703(11)) but fails on OOD evaluation, with scores dropping to the random-guess baseline (0.493(11) ``Real'' and 0.524(9) ``Synth''). 
Conversely, datasets that generalize well exhibit the opposite pattern. 
Our \dataset dataset, for example, yields a modest ID accuracy (0.590(3)) but achieves the highest OOD performance across all experiments, scoring 0.881(26) on ``Real'' and 0.785(7) on ``Synth'' samples.
This pattern of low ID and high OOD performance is also seen in CVEfixes (0.518(2) ID → 0.837(8) ``Real'') and CleanVul (0.569(9) ID → 0.790(21) ``Real''). Other datasets show mixed results: DiverseVul and PrimeVul generalize moderately, primarily on the ``Synth'' portion of \benchmark, while SafeCoder shows moderate performance on the ``Real'' portion. VulnPatchPairs performs poorly across all metrics, with both ID (0.562(9)) and OOD (0.543(15) ``Real'') scores remaining just above 0.5.
This demonstrates a clear disconnect between ID testing, which appears to reward overfitting (e.g., BigVul), and OOD generalization. Datasets like \dataset, CVEfixes, and CleanVul produce models that generalize effectively, despite their modest ID scores, highlighting their utility for training models on real-world tasks.

\begin{table}[b]
\centering
\vspace{-2mm}
\caption{Qwen2.5-Coder-1.5B Accuracy on \benchmark (C/C++) after training on various datasets.}
\vspace{-2mm}
\label{tb:model_performance_on_c}
\resizebox{0.65\columnwidth}{!}{%
\begin{tabular}{l|c|c}
\toprule
\textbf{Dataset} & \textbf{ID} & \textbf{\benchmark in C/C++} \\
\midrule
\textbf{BigVul} & \cellcolor{color-f}\textbf{\num{0.703(11)}} & \num{0.517(7)} \\
\textbf{CVEfixes} & \num{0.518(2)} & \cellcolor{color-f}\num{0.746(44)} \\
\textbf{CleanVul} & \num{0.569(9)} & \cellcolor{color-f}\num{0.781(16)} \\
\textbf{DiverseVul} & \num{0.538(4)} & \cellcolor{color-f}\num{0.763(11)} \\
\textbf{\dataset} & \num{0.590(3)} & \cellcolor{color-f}\textbf{\num{0.782(14)}} \\
\textbf{PrimeVul} & \num{0.539(6)} & \cellcolor{color-f}\num{0.740(6)} \\
\textbf{SafeCoder} & \num{0.531(4)} & \num{0.602(19)} \\
\textbf{VulnPatchPair} & \num{0.562(9)} & \num{0.563(55)} \\
\bottomrule
\end{tabular}%
}
\vspace{-3mm}
\end{table}

\myparagraph{Consistency of Findings on C/C++ Specific Benchmarking}
Furthermore, because several training datasets (e.g., BigVul, DiverseVul) are predominantly C/C++, we conducted a controlled experiment to ensure these findings are not confounded by language-specific factors. We evaluated the Qwen2.5-Coder-1.5B model against only the C/C++ samples from \benchmark. The results, presented in \Cref{tb:model_performance_on_c}, confirm the same trend. For instance, BigVul retains a large gap (0.703(11) ID → 0.517(7) OOD), while \dataset (0.590(3) ID → 0.782(14) OOD) and CleanVul (0.569(9) ID → 0.781(16) OOD) again show strong generalization from low ID scores. This consistency reinforces that the observed limitations stem from fundamental data quality issues, not language mismatches.

\begin{table}[b]
\centering
\vspace{-2mm}
\caption{Performance of models evaluated on \benchmark and on ReVeal and Real-Vul.}
\vspace{-2mm}
\label{tb:external_performance}
\resizebox{0.48\textwidth}{!}{%
\begin{tabular}{l|ccc|ccc}
\toprule
\multirow{2}{*}{\textbf{Model}} & 
\multicolumn{3}{c|}{\textbf{Trained on ReVeal}} & 
\multicolumn{3}{c}{\textbf{Trained on Real-Vul}} \\
\cmidrule(lr){2-4}
\cmidrule(lr){5-7}
& \textbf{ID} & \textbf{Real} & \textbf{Synth} & 
\textbf{ID} & \textbf{Real} & \textbf{Synth} \\
\midrule

\textbf{CodeBERT} & 
\cellcolor{color-f}\num{0.762(8)} & \cellcolor{color-e}\num{0.498(2)} & \cellcolor{color-e}\num{0.481(2)} &
\cellcolor{color-b}\num{0.963(1)} & \cellcolor{color-e}\num{0.497(15)} & \cellcolor{color-e}\num{0.487(12)} \\

\textbf{GraphCodeBERT} & 
\cellcolor{color-f}\num{0.781(6)} & \textbf{\num{0.500(7)}} & \cellcolor{color-e}\textbf{\num{0.493(5)}} &
\cellcolor{color-b}\num{0.966(4)} & \cellcolor{color-e}\num{0.494(2)} & \cellcolor{color-e}\num{0.492(5)} \\

\textbf{UniXcoder} & 
\cellcolor{color-f}\num{0.789(10)} & \cellcolor{color-e}\textbf{\num{0.493(16)}} & \cellcolor{color-e}\num{0.412(3)} &
\cellcolor{color-b}\num{0.968(1)} & \cellcolor{color-e}\num{0.497(12)} & \num{0.506(18)} \\

\textbf{Llama-3.2-3B} & 
\cellcolor{color-b}\textbf{\num{0.800(22)}} & \cellcolor{color-e}\num{0.496(8)} & \cellcolor{color-e}\num{0.424(15)} &
\cellcolor{color-b}\textbf{\num{0.969(2)}} & \textbf{\num{0.504(3)}} & \textbf{\num{0.511(7)}} \\

\textbf{Qwen2.5-Coder-1.5B} &
\cellcolor{color-f}\num{0.750(11)} & \cellcolor{color-e}\num{0.491(8)} & \cellcolor{color-e}\num{0.466(26)} &
\cellcolor{color-b}\num{0.967(5)} & \num{0.500(9)} & \num{0.500(14)} \\

\bottomrule
\end{tabular}%
}
\vspace{-3mm}
\end{table}

\myparagraph{The Pitfall of ``Vulnerability-Like'' Code Detection} To further probe dataset-induced biases, we evaluated generalization on two additional external datasets, ReVeal and Real-Vul. As these datasets originally contained much more benign code than vulnerable code, we created balanced versions for a fair comparison by down-sampling the benign class to match the number of vulnerable samples. As shown in \Cref{tb:external_performance}, models trained on these datasets achieve excellent In-Distribution (ID) performance. Llama-3.2-3B, for instance, achieves an ID accuracy of 0.800(22) on ReVeal, and all models score above 0.960 on Real-Vul. This high ID performance is likely attributable to their construction: these datasets pair vulnerable code with benign samples from the same repository, not with their corresponding fixed versions. This task of differentiating vulnerable code from general benign code might be much simpler. 
However, when evaluated on the OOD setting (\benchmark), the performance of these models drops to near the random-guessing baseline. This provides an important insight: models trained on datasets that use general benign code as negative samples (like ReVeal and Real-Vul), rather than vulnerability-fix pairs, may only learn to identify \emph{vulnerability-like} code. They are struggling to differentiate a vulnerability from its corresponding fix. This suggests the model has not learned the precise, semantic nature of the vulnerability, which is a critical limitation for future vulnerability detection studies to overcome.

\myparagraph{Zero-Shot and Few-Shot LLM Baseline Performance}
To contextualize the performance of our fine-tuned models, we evaluated powerful LLM baselines (GPT4.1 and Claude-3.7-Sonnet) on \benchmark using prompt-based methods, as shown in \Cref{tb:zero_shot}. We tested Direct (zero-shot), CoT (zero-shot Chain-of-Thought), and ICL (three-shot In-Context Learning) strategies. Overall, these advanced baselines perform modestly, with most accuracies falling between 0.5 and 0.7, indicating that accurately identifying vulnerabilities in vulnerability-fix pair datasets is challenging for zero-shot or few-shot learners. Performance was notably higher on Real-Vul, where Claude-3.7-Sonnet (Direct) achieved 0.700(1), suggesting the task in that dataset (differentiating vulnerable from general benign code) is simpler for LLMs.

\begin{table}[tbp]
\centering
\caption{GPT4.1 and Claude-3.7-Sonnet using zero-shot (Direct, CoT) and few-shot (ICL) prompting strategies.}
\vspace{-2mm}
\label{tb:zero_shot}
\resizebox{\columnwidth}{!}{%
\begin{tabular}{l|ccc|ccc}
\toprule
\multirow{2}{*}{\textbf{Dataset}} & \multicolumn{3}{c|}{\textbf{GPT4.1}} & \multicolumn{3}{c}{\textbf{Claude-3.7-Sonnet}} \\
\cmidrule(lr){2-4}
\cmidrule(lr){5-7}
 & \textbf{Direct} & \textbf{CoT} & \textbf{ICL} & \textbf{Direct} & \textbf{CoT} & \textbf{ICL} \\
\midrule
\textbf{BigVul} & \num{0.504(4)} & \textbf{\num{0.511(11)}} & \num{0.504(2)} & \num{0.520(5)} & \textbf{\num{0.530(8)}} & \num{0.525(11)} \\
\textbf{CVEfixes} & \num{0.502(2)} & \num{0.510(6)} & \textbf{\num{0.525(14)}} & \num{0.505(1)} & \textbf{\num{0.511(4)}} & \num{0.509(7)} \\
\textbf{CleanVul} & \num{0.529(2)} & \textbf{\num{0.535(6)}} & \num{0.529(4)} & \num{0.526(4)} & \num{0.529(3)} & \textbf{\num{0.527(6)}} \\
\textbf{DiverseVul} & \textbf{\num{0.517(7)}} & \num{0.508(16)} & \num{0.513(8)} & \num{0.517(10)} & \textbf{\num{0.523(3)}} & \num{0.519(17)} \\
\textbf{PrimeVul} & \num{0.508(3)} & \textbf{\num{0.528(11)}} & \num{0.511(4)} & \num{0.511(1)} & \num{0.516(2)} & \textbf{\num{0.520(8)}} \\
\textbf{SafeCoder} & \num{0.572(3)} & \textbf{\num{0.586(7)}} & \num{0.575(23)} & \textbf{\num{0.605(4)}} & \num{0.571(8)} & \num{0.586(28)} \\
\textbf{VulnPatchPair} & \num{0.509(3)} & \num{0.500(7)} & \textbf{\num{0.512(7)}} & \num{0.503(3)} & \textbf{\num{0.512(7)}} & \num{0.503(5)} \\
\midrule
\textbf{\dataset} & \num{0.518(0)} & \num{0.515(8)} & \textbf{\num{0.521(6)}} & \num{0.503(3)} & \textbf{\num{0.512(7)}} & \num{0.503(5)} \\
\textbf{\benchmark Real} & \num{0.623(4)} & \num{0.626(7)} & \textbf{\num{0.659(18)}} & \num{0.639(0)} & \num{0.586(2)} & \textbf{\num{0.654(27)}} \\
\textbf{\benchmark Synth} & \num{0.597(2)} & \num{0.634(8)} & \textbf{\num{0.669(18)}} & \num{0.520(5)} & \textbf{\num{0.530(8)}} & \num{0.525(11)} \\
\midrule
\textbf{Real-Vul} & \num{0.676(6)} & \num{0.629(2)} & \textbf{\num{0.691(29)}} & \cellcolor{color-f}\textbf{\num{0.700(1)}} & \num{0.630(4)} & \num{0.668(11)} \\
\textbf{ReVeal} & \num{0.564(3)} & \num{0.546(10)} & \textbf{\num{0.573(8)}} & \textbf{\num{0.580(4)}} & \num{0.516(6)} & \num{0.566(3)} \\
\bottomrule
\end{tabular}%
}
\vspace{-3mm}
\end{table}

\input{figures/comparison_2}
\input{figures/comparison_1}

\takeaway{ {\bf Answer to RQ1:} \textbf{In-Distribution (ID) performance is a poor indicator of Out-of-Distribution (OOD) generalization.} Our results show that ID performance has almost no clear relation to OOD performance on \benchmark. For example, \textbf{BigVul} achieves a high 0.703 ID accuracy but fails on OOD evaluation (0.493 on ``Real''). Conversely, \textbf{\dataset} yields a modest ID score (0.590) but achieves the highest OOD performance (0.881 on ``Real''), demonstrating effective generalization. }

\vspace{-1mm}
\subsection{RQ2: Effect of Training Data Across CWEs}

To further investigate the performance differences observed in RQ1, we compare model performance across the Top 25 Most Dangerous CWE categories to evaluate how the choice of training dataset affects generalization at a per-CWE level. The results are detailed in \Cref{fig:cwe_performance_2} and \Cref{fig:cwe_performance_1}.

\myparagraph{Weak Datasets Fail Across CWEs}
The per-CWE analysis confirms our findings from RQ1. For datasets that performed poorly on OOD evaluation, such as BigVul and VulnPatchPairs, \Cref{fig:cwe_performance_1} and \Cref{fig:cwe_performance_2} show that this is not an averaging effect. Rather, models trained on these datasets perform at or near the 0.5 random-guess baseline for almost every single CWE category. This indicates that the models failed to learn any generalizable vulnerability patterns, likely due to overfitting on dataset-specific artifacts.

\myparagraph{Effective Datasets Show ``Spiky'' CWE-Specific Biases}
Conversely, datasets that achieved high OOD performance in RQ1, such as CleanVul and CVEfixes, show a different profile. Their models learn genuine vulnerability features, but their expertise is spiky and reflects the specific emphasis of each dataset. For example, the model trained on CleanVul is highly effective at detecting CWE-862 (0.79 accuracy), whereas the CVEfixes-trained model struggles with the same category (0.62 accuracy). This pattern suggests that while both datasets are effective, they are not interchangeable and have different strengths. The performance profiles of \dataset show the most consistent high performance across the broadest range of CWEs, correlating with its high OOD score in RQ1.

\myparagraph{Data Sparsity as a Bottleneck for Rare CWEs}
It is worth noting that for CWE-798 (Hard-coded Credentials), all models perform poorly, regardless of the training dataset. This result is highly consistent with our initial data analysis (referenced in \Cref{fig:cwe25}), which identified CWE-798 as the least common category in the combined dataset. This strongly suggests that a very small number of samples for a specific vulnerability type has a negative impact on the model's ability to learn its patterns, highlighting an important challenge for future dataset curation.

\takeaway{
{\bf Answer to RQ2:}
\textbf{The choice of dataset decisively shapes per-CWE performance.} Datasets like \textbf{BigVul} and \textbf{VulnPatchPairs} show low performance across the majority of CWEs, suggesting that features learned from them do not generalize well. In contrast, \textbf{\dataset} provides the most consistent high performance, with a per-CWE range of 0.59 to 0.93. We also find that \textbf{data sparsity is a key bottleneck}, as all models perform poorly on the least common category, CWE-798.
}

\vspace{-1mm}
\subsection{RQ3: Impact of Synthesizing Training Data}

Synthesizing vulnerability examples with LLMs offers a promising way to augment real-world vulnerability datasets, potentially addressing data scarcity for underrepresented CWEs. To explore this hypothesis, we augmented the \dataset training set with synthesized vulnerabilities. 
Specifically, we added 100 new vulnerable samples and their corresponding fixes for each of the 25 CWE types using our \rvg pipeline (\Cref{sec:rvg}).
There is no duplication between these new synthesized samples and our \benchmark evaluation set, ensuring evaluation integrity.
We then compared the performance of Qwen2.5-Coder-1.5B trained on the original \dataset versus this augmented dataset. As shown in \Cref{fig:titanvul}, the inclusion of synthesized data improves performance. The model's accuracy on the ``Real'' portion of \benchmark improves from 0.881(26) to 0.932(7) (a 5.8\% increase), and its accuracy on the ``Synth'' portion rises from 0.785(7) to 0.888(5) (a 13.1\% increase). This provides a key insight: adding synthetic training data not only improves performance on other synthetic data but also measurably improves generalization on unseen, real-world vulnerabilities.

\begin{figure}[htb]
\begin{minipage}{0.5\textwidth}
\scalebox{0.38}{
\begin{tikzpicture}
    \centering
    \begin{axis}[
        height=8.5cm, width=22.5cm,
        /pgf/bar width=0.26cm,
        xmin=0.95, xmax=19.05,
        axis x line*=bottom, axis y line*=left, enlarge x limits=true,
        xtick={0,1,2,3,4,5,6,7,8,9,10,11,12,13,14,15,16,17,18,19,20},
        xticklabel style={yshift=-1mm, font=\LARGE, align=center, rotate=45, anchor=east},
        ybar=3.8pt, clip=false,
        ymin=0, ymax=1, 
        ytick={0, 0.2, 0.4, 0.6, 0.8, 1},
        yticklabels={0, 0.2, 0.4, 0.6, 0.8, 1},
        ymajorgrids, major grid style={draw=black!20}, tick align=inside,
        yticklabel style={font=\huge}, tickwidth=0pt,
        y axis line style={opacity=0},
        ylabel={\huge Accuracy: \dataset Without vs. With Synthesized Data},
        y label style={at={(0.32, 1.15)}, rotate=-90},
        xticklabels={
            125, 190, 200, 22, 269, 306, 352, 400, 416, 434, 476, 502, 78, 787, 79, 798, 862, 863, 89, 918, 94
        },
    ]

    \addplot [draw=light-gray, line width=0.7pt, fill=color-f, error bars/.cd, y dir=both, y explicit, error bar style={draw=black}] coordinates {
        (0, 0.90) +- (0, 0.01)
        (1, 0.85) +- (0, 0.01)
        (2, 0.76) +- (0, 0.02)
        (3, 0.91) +- (0, 0.01)
        (4, 0.73) +- (0, 0.04)
        (5, 0.76) +- (0, 0.03)
        (6, 0.73) +- (0, 0.02)
        (7, 0.81) +- (0, 0.07)
        (8, 0.65) +- (0, 0.03)
        (9, 0.82) +- (0, 0.03)
        (10, 0.86) +- (0, 0.04)
        (11, 0.76) +- (0, 0.05)
        (12, 0.86) +- (0, 0.03)
        (13, 0.86) +- (0, 0.01)
        (14, 0.90) +- (0, 0.03)
        (15, 0.59) +- (0, 0.02)
        (16, 0.81) +- (0, 0.05)
        (17, 0.73) +- (0, 0.02)
        (18, 0.93) +- (0, 0.00)
        (19, 0.78) +- (0, 0.04)
        (20, 0.86) +- (0, 0.02)
    };
    
    \addplot [draw=light-gray, line width=0.7pt, fill=color-b, error bars/.cd, y dir=both, y explicit, error bar style={draw=black}] coordinates {
        (0, 0.98) +- (0, 0.02)
        (1, 0.88) +- (0, 0.01)
        (2, 0.80) +- (0, 0.01)
        (3, 0.79) +- (0, 0.02)
        (4, 0.88) +- (0, 0.03)
        (5, 0.83) +- (0, 0.04)
        (6, 0.90) +- (0, 0.01)
        (7, 0.97) +- (0, 0.02)
        (8, 0.74) +- (0, 0.01)
        (9, 0.94) +- (0, 0.01)
        (10, 0.90) +- (0, 0.01)
        (11, 0.92) +- (0, 0.03)
        (12, 0.93) +- (0, 0.01)
        (13, 0.94) +- (0, 0.01)
        (14, 0.93) +- (0, 0.01)
        (15, 0.86) +- (0, 0.02)
        (16, 0.92) +- (0, 0.02)
        (17, 0.84) +- (0, 0.02)
        (18, 0.99) +- (0, 0.01)
        (19, 0.94) +- (0, 0.02)
        (20, 0.94) +- (0, 0.01)
    }; 
    \end{axis}
\end{tikzpicture}
}
\end{minipage}
\vspace{-0mm}

\begin{minipage}{0.5\textwidth}
\centering
\label{tab:model_performance}
\resizebox{0.8\textwidth}{!}{%
\begin{tabular}{l|cc}
\toprule
\textbf{Dataset} & \textbf{Real} & \textbf{Synth} \\
\midrule
\dataset without Synthesized Data & \cellcolor{color-b}\textbf{\num{0.881(26)}} & \cellcolor{color-f}\textbf{\num{0.785(7)}} \\
\dataset with Synthesized Data & \cellcolor{color-b}\textbf{0.932(7)} (5.8\%$\uparrow$) & \cellcolor{color-b}\textbf{0.888(5)} (13.1\%$\uparrow$) \\
\bottomrule
\end{tabular}
}
\end{minipage}
\caption{Performance on \dataset without synthesized data vs. with synthesized data.}
\label{fig:titanvul}
\vspace{-5mm}
\end{figure}

\myparagraph{Targeted Improvement for Under-Representative Weaknesses}
The benefits of data synthesis are most pronounced for weaknesses that are rare in the original dataset. A clear example is CWE-798 (Hard-Coded Credentials), for which data was scarce. This data scarcity limited the baseline model to 0.587(20) accuracy; however, after augmentation, its accuracy surged to 0.863(15), which is one of the most substantial gains observed in the per-CWE analysis. Conversely, for vulnerabilities where the baseline model was already highly proficient (e.g., CWE-125), the gains were more modest. This demonstrates that data synthesis is a powerful tool for compensating for data scarcity while still offering incremental benefits for well-represented classes.

\takeaway{
\textbf{Answer to RQ3:} 
\textbf{Augmenting \dataset with synthesized data improves OOD performance on both real-world and synthetic data.} Accuracy on \benchmark's ``Real'' portion increases by 5.8\% (0.881 → 0.932), and on the ``Synth'' portion by 13.1\% (0.785 → 0.888). \textbf{Gains are especially notable for underrepresented weaknesses}, such as \textbf{CWE-798}, where accuracy rises from 0.587(20) to 0.863(15).
}

\section{Discussion}
\label{sec:discussion}

\myparagraph{The Deception of In-Distribution (ID) Evaluation} Our results challenge the validity of ID evaluation as a meaningful performance metric. We find that ID accuracy does not reliably indicate OOD performance on \benchmark. This is most evident with BigVul: the Qwen2.5-Coder-1.5B model achieves a high 0.703(11) ID accuracy, but its OOD performance on \benchmark's     ``Real'' drops to 0.493(11), close to random guessing. This suggests severe overfitting to dataset-specific artifacts. This disconnect is further highlighted when comparing \dataset and VulnPatchPairs. Despite having similar modest ID scores (0.590(3) and 0.562(9), respectively), their generalization performance diverges dramatically. \dataset achieves the highest OOD performance (0.881(26)), while VulnPatchPairs remains low (0.543(15)). This demonstrates that ID performance is a misleading indicator of generalization, underscoring the necessity of high-quality, independent benchmarks like \benchmark to assess a model's true detection capabilities.

\myparagraph{Beyond Validity: The Critical Role of Negative Samples}
Our results show two failure modes for generalization. First, high-noise datasets like BigVul (25.0\% \emph{validity}) and VulnPatchPairs (36.0\% \emph{validity}) fail to generalize, with OOD (``Real'') accuracies near the random-guess. Second, construction methodology is also critical. Models trained on Real-Vul and ReVeal achieve high ID scores (e.g., > 0.960 on Real-Vul) but also fail on \benchmark's OOD evaluation. This is likely because they pair vulnerable code with general benign code, not with vulnerability-fix pairs. This indicates that these models learn to find \emph{vulnerability-like} code but are surprisingly unable to differentiate a vulnerability from its corresponding fix when evaluated on \benchmark. This highlights the critical and nuanced impact of negative sample choice on dataset construction.

\myparagraph{Threats to Validity}
We acknowledge several potential validity concerns and outline mitigation steps. First, the validity of our \benchmark is a key consideration. To address this, we implemented a multi-stage validation process that included both automated filtering and manual review. Our additional manual assessment confirms a high \emph{correctness} rate of 92\%, demonstrating that \benchmark is accurate for evaluating vulnerability detection models. This approach to dataset validation is consistent with standards adopted in related empirical software engineering studies~\cite{ray2014large}. 
In addition, \benchmark's reliance on synthetic augmentation (\rvg) poses a potential threat, which we mitigate through rigorous manual review.
Moreover, our function-level detection focus, while a known limitation, is widely accepted in vulnerability detection research~\cite{wu2022vulcnn,liu2024vuldetectbench,risse2025top}; both \benchmark and \dataset have been specifically constructed and validated for function-level granularity, clearly delimiting the scope and applicability of our findings. 
Finally, observed effects may reflect LLM capabilities or prompting choices; while our pipeline is empirically effective, we do not isolate strategy-specific benefits through explicit baselines or ablations.
Taken together, these mitigation strategies and acknowledged limitations align our evaluation with best practices in the field.

\section{Related Work}
\label{sec:related_work}

\myparagraph{Vulnerability Datasets}
Early datasets mined from commits, like BigVul \cite{fan2020ac}, suffer from significant noise, with \emph{validity rates} as low as 25.0\% \cite{ding2024vulnerability}. Subsequent datasets made trade-offs: CVEfixes \cite{bhandari2021cvefixes} offers precise CVE mapping but has limited scope; PrimeVul \cite{ding2024vulnerability} achieves 86.0\% \emph{validity} by focusing on single-function commits, potentially sacrificing realism; and DiverseVul \cite{chen2023diversevul} prioritizes language diversity at the cost of \emph{validity} (60.0\%). CleanVul \cite{li2024cleanvul} advanced the field by using LLMs to filter noisy commits, achieving 90.6\% \emph{validity}. Our work builds on these efforts. We construct \dataset by applying a rigorous, multi-agent LLM verification pipeline to a large, aggregated corpus, uniquely combining scale and high quality. Furthermore, since self-evaluation overestimates real-world performance \cite{risse2025top}, an independent benchmark is needed. We introduce \benchmark, the first manually-verified, balanced benchmark providing comprehensive coverage of the MITRE Top 25 Most Dangerous CWEs \cite{mitre2024cwe}. These contributions enable more reliable evaluation and foster the development of models with true generalizability.

\myparagraph{LLMs for Software Security}
Recent work has explored the use of LLMs across a broad range of software security tasks, including vulnerability detection~\cite{chakraborty2021deep,ding2024vulnerability,gao2023far}, code clone analysis~\cite{dou2023towards,zhang2024assessing,bui2025vulcoco}, dataset construction~\cite{li2024cleanvul,nguyen2025patchseeker,chen2025forge}, automated vulnerability repair~\cite{islam2024llm,kulsum2024case,yang2025semantics}, and secure code generation~\cite{he2023large,li2025gensiac,chen2025secureagentbench}. 
However, the use of LLMs for vulnerability synthesis remains limited. Existing approaches primarily rely on programmatic injection or neural code editing~\cite{dolan2016lava,nong2022generating,nong2023vulgen}, which often lack realistic development context. 
In contrast, our \rvg leverages LLMs to synthesize realistic, context-aware vulnerability/fix pairs across different CWEs.

\vspace{2mm}
\section{Conclusion and Future Work}
\label{sec:conclusion}

We present \benchmark, a manually-verified and balanced benchmark for the MITRE Top 25 Most Dangerous CWEs \cite{mitre2024cwe}, enabling reliable evaluation of model generalization. We also construct \dataset, a large-scale, high-quality dataset (38,548 vulnerable functions, 94\% \emph{validity rate}), curated via a novel multi-agent LLM pipeline, and propose \rvg for synthesizing realistic, context-aware vulnerability data to tackle data scarcity.
Our experiments show that In-Distribution (ID) performance (testing on the same dataset) is misleading and does not reliably indicate Out-of-Distribution (OOD) generalization. 
For example, models trained on BigVul achieve high ID accuracy (0.703) but fail on \benchmark's real-world portion (0.493).
Conversely, models trained on our \dataset achieve the highest OOD performance (0.881) on \benchmark's real-world portion, despite a modest ID score (0.590).
Augmenting \dataset with \rvg-generated data further enhances this OOD performance, improving accuracy on real-world data by 5.8\% (to 0.932).
In future work, we plan to extend our resources to cover a broader range of CWEs, support inter-procedural vulnerability analysis, and assess the applicability of our benchmarks and datasets to large-scale, real-world industrial codebases.

\vspace{-1mm}
\begin{acks}
This research / project is supported by the National Research Foundation, Singapore, and the Smart Nation Group under the Smart Nation Group's Translational R\&D Grant (Award No. TRANS2023-TGC02). Any opinions, findings and conclusions or recommendations expressed in this material are those of the author(s) and do not reflect the views of National Research Foundation, Singapore or the Smart Nation Group.
\end{acks}

\clearpage  %
\balance
\bibliography{bibliography}
\bibliographystyle{ACM-Reference-Format}

\end{document}